\documentclass[twocolumn,pra,aps,showpacs]{revtex4}
\usepackage{epsfig}
\usepackage[english]{babel}
\usepackage{latexsym}
\usepackage{graphics}
\usepackage{subfigure}

\begin{document}
\title{Preparation of polarization entangled mixed states of two photons}
\author{Chuanwei Zhang}
\begin{abstract}
We propose a scheme for preparing arbitrary two photons polarization entangled mixed
states via controlled location decoherence. The scheme uses
only linear optical devices and single-mode optical fibers, and may be
feasible in experiment within current optical technology.
\end{abstract}
\address{Department of Physics and Center for Nonlinear Dynamics, The University of Texas, Austin, Texas
78712-1081}
\pacs{03.67.-a, 42.25.Ja, 03.65.Ud, 89.70.+c}
\maketitle

Entanglement has played a crucial role for many applications of quantum
information, such as quantum teleportation \cite{Ben93}, superdense coding 
\cite{Ben92}, quantum error correction \cite{Got}, etc. To function
optimally these applications requires maximal pure entanglement. However,
unwanted coupling to the environment causes decoherence of quantum systems
and yields mixed state entanglement. Therefore entanglement concentration 
\cite{Kwiat01,Pan01,Munro01,Thew01,Bouda02} and various applications of mixed state
entanglement \cite{Cleve99,Bowen,Verstraete} are very important and have been
investigated by many authors. Experimentally, a special mixed state,
``decoherence-free subspace'' \cite{Kwiat00}, has been demonstrated and 
optical Werner states have been prepared \cite{GCG}.

In this paper, we propose a scheme for preparing arbitrary
polarization-entangled mixed states of two photons via controlled
decoherence. In the experiment of demonstrating decoherence-free subspace,
decoherence was imposed by coupling polarization modes to frequency modes of
photons. Here we introduce decoherence by entangling polarization modes with
location modes of photons, where location modes are finally traced out
by mixing them with appropriate path length differences and detecting coincidences
independent on the emission of photon pair from photon source. 

Consider a two-qubit mixed state $\rho $ of quantum system $AB$, it can be
represented as \cite{Woot98} 
\begin{equation}
\rho =\sum_{i=1}^4p_i\left| \psi _i\right\rangle _{AB}\left\langle \psi
_i\right|,  \label{1}
\end{equation}
where $0\leq p_i\leq 1$, $\sum_{i=1}^4p_i=1$, and $p_i\geq p_j\ $for $i\leq j
$. $\left| \psi _i\right\rangle _{AB}$ are two-qubit pure states with same
entanglement of formation (EOF) as $\rho $. Therefore $\left| \psi
_i\right\rangle $ are same up to some local unitary operations \cite{Niel99}%
, i.e. $\left| \psi _i\right\rangle =U_i\otimes V_i\left| \Phi \right\rangle
$, where $U_i$ and $V_i$ are local unitary operations and $\left| \Phi
\right\rangle =\cos \theta \left| 00\right\rangle +\sin \theta \left|
11\right\rangle $ with $0\leq \theta \leq \pi /4$.

The experimental arrangement for our scheme is described as Fig.1. First,
spontaneous parametric down-conversion in two adjacent $\beta $-barium
borate (BBO) crystals produces initial two photons polarization-entangled
pure state $\left| \Phi \right\rangle _{AB}=\cos \theta \left|
HH\right\rangle +\sin \theta \left| VV\right\rangle $ \cite{Kwiat95,White99},
where $\left| H\right\rangle $ and $\left| V\right\rangle $ are horizontal
and vertical polarizations respectively. Then six beam splitters with
variable transmission coefficients (VBS) couple the initial
polarization state $\left| \Phi \right\rangle _{AB}$ to location modes and
each photon has four possible optical paths $i_{A(B)}$. At
each path, single-qubit polarization rotations (SPR) perform local unitary
operations $U_i, V_i$ on the polarization mode of each photon and transform initial
entangled state $\left| \Phi \right\rangle _{AB}$ to different $\left| \psi
_i\right\rangle _{AB}$. The four paths of each photon mix on couplers $%
G_{A(B)}$ and become one through single-mode optical fibres \cite{Tittel98}. 
In experiment, lossless mixture process can be realized by using a fast switch to combine
different modes, in addition to a pulsed pump laser. However, this method is
not very practical although it can be implemented in principle. A practical replacement can be 
a passive coupler without switch although 
it will decrease the optimal success probability due to losses in the coupling.    
Denote $L_i^{A(B)}$ as the optical path lengths of paths $i_{A(B)}$ (from the
BBO crystal to coupler $G_{A(B)}$). If $L_i^{A(B)}$ satisfy $L_i^A=$ $%
L_i^B$ and $\Delta _{ij}^{A(B)}=\left| L_i^{A(B)}-L_j^{A(B)}\right| \gg
l_{coh}$ for different $i$ and $j$, the location modes will be traced out and
we obtain a mixed state, where $l_{coh}$ is the single-photon coherence
length. 

\begin{figure}[!t]
\begin{center}
\vspace*{-0.5cm} \resizebox *{9cm}{6.5cm}{\includegraphics*{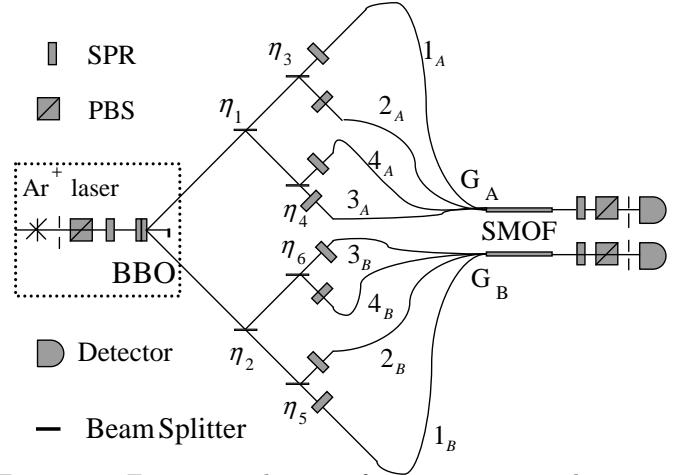}}
\end{center}
\par
\vspace*{-1.0cm}
\caption{ Experimental set-up for preparing an arbitrary
polarization-entangled mixed state of two photons.}
\label{fig:1}
\end{figure}

The mixed state still contains information of the location modes since
photons from different paths $i$ will arrive at the detector at different time.
Therefore the mixed state is composed by two discrete
subspace state: two photons $A$, $B$ arriving at same time and at different
time. If the time window of the coincidence counter is small enough,
only photons from paths with same lengths ($i_A $ and $i_B$)
contribute the counts and the polarization states of photons are
reduced to the subspace state with same arrival time. In this subspace, the state
is just the two-qubit mixed state $\rho $. Therefore the final SPR in each arm, along with PBS, enable analysis of the
polarization correlations in any basis, allowing tomographic reconstruction
of the density matrix \cite{Kwiat00,White99,Kwiat01}.

In our scheme, we require that the coherence length of pump laser is much smaller 
than path length difference $\Delta _{ij}$ in order to avoid two-photons interference. 
In \cite{Tittel98}, two-photons interference
can be used to realize a Franson-type test of Bell inequalities, where the path length difference is 2 orders of magnitude
smaller than the coherence length of the pump laser. In our scheme, 
we require that the path length difference is larger than both single photon and pump 
laser coherence lengths to avoid both single and two photons interference. In this case, 
the travel time of a photon pair from laser to detector 
enables to resolve the different paths in principle. It is therefore important to post-select
only photons arriving in coincidence but not to resolve the travel 
time from emission to detection in order to trace out all location modes.
  
The VBS in the scheme can be implemented using a one-order Mach-Zehnder
interferometer \cite{zhang01,Reck94} and it transforms location modes in the
following way
\begin{equation}
\left| a\right\rangle _{initial}\rightarrow \sqrt{\eta _i}\left|
a\right\rangle _{final}+\sqrt{1-\eta _i}\left| b\right\rangle _{final},
\label{2}
\end{equation}
where $\left| a\right\rangle $ and $\left| b\right\rangle $ are the location
modes and $\sqrt{\eta _i}$ are the transmission coefficients. The $%
SPR_i^{A(B)}$ at paths $i_{A(B)}$ can be constructed with wave plate
sequences $\left\{ \text{QWP, HWP, QWP}\right\} $ \cite{zhang01} and they
perform unitary operation $U_i^A$($V_i^B$). The coupler $G_{A(B)}$ introduces 
decoherence, which yields mixed state 
\begin{equation}
\rho _0=\sum_{i,j=1}^4p_{ij}U_i\otimes V_j\left| \Phi \right\rangle
\left\langle \Phi \right| U_i^{\dagger }\otimes V_j^{\dagger }\text{,}
\label{3}
\end{equation}
where $p_{ij}$ is the combined probability with photons $A$ and $B$ at
paths $\left| i\right\rangle _A$, $\left| j\right\rangle _B$ respectively.
If the time window of the coincidence counter $T$ satisfies 
$T<\Delta _{ij}/c$ ($c$ is the velocity of the light), only
photons with location modes $\left| i\right\rangle _A\left| i\right\rangle _B$
are registered by the coincidence counter, and density matrix $%
\rho _0$ is reduced to 
\begin{equation}
\rho =\frac 1F\sum_{i=1}^4p_{ii}\left| \psi _i\right\rangle \left\langle
\psi _i\right|,  \label{4}
\end{equation}
where $F=\sum_{i=1}^4p_{ii}$ is the
successful probability of generating the mixed state $\rho $.

\begin{figure}[!t]
\begin{center}
\vspace*{-0.5cm} \resizebox *{9cm}{4cm}{\includegraphics*{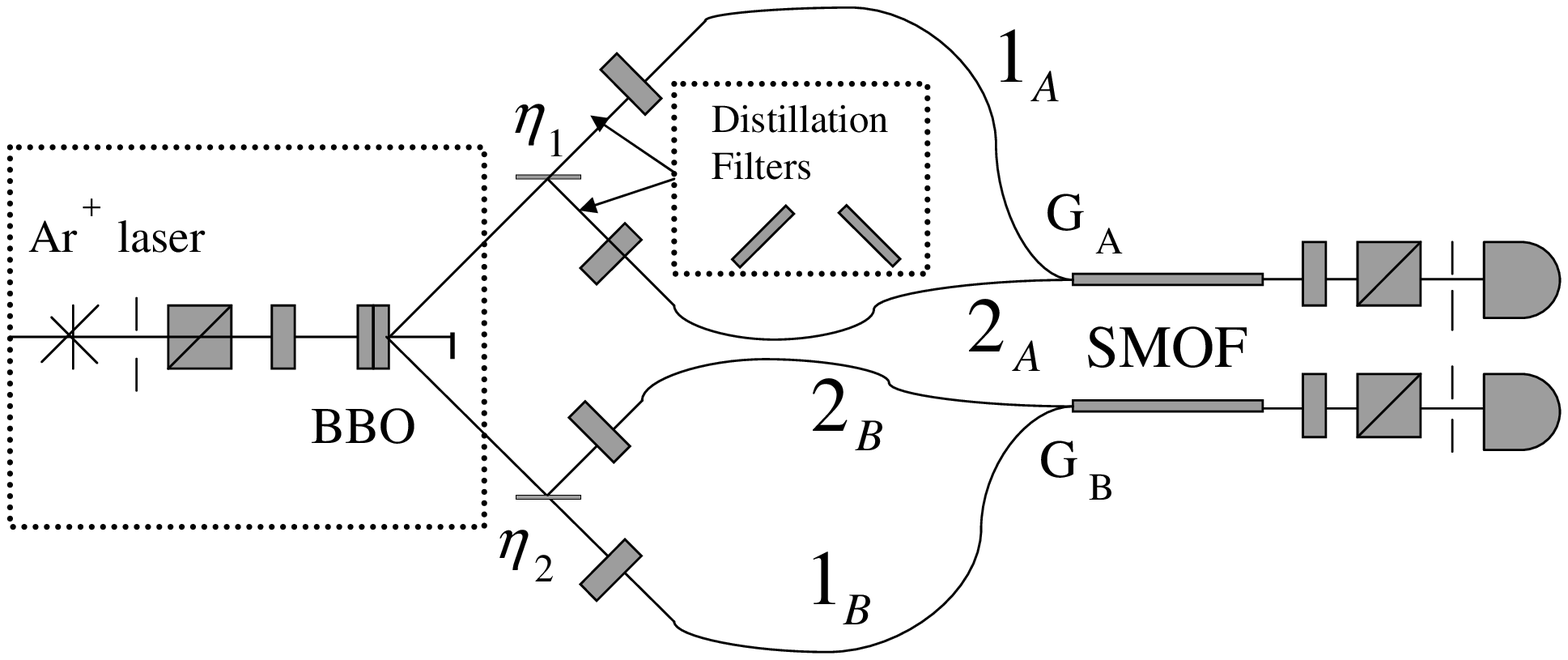}}
\end{center}
\par
\vspace*{-1.0cm}
\caption{ Experimental set-up used to generate mixed states $\rho
=p\left| \psi \right\rangle \left\langle \psi \right| +\left( 1-p\right)
\left| \phi \right\rangle \left\langle \phi \right| $.}
\label{fig:2}
\end{figure}
Denote $p_i=p_{ii}/F$, the remaining problem is to find the optimal
successful probability $F$. For a given density matrix of a mixed state,
there are several choices of $\left| \psi_i \right\rangle $ (to be realized using different local operation and different 
initial entangled states) and thus also different beamsplitter settings. The best choice is the one where the success probability 
is maximized. From Fig.1, we find $p_{11}=\eta _1\eta _2\eta
_3\eta _5$, $p_{22}=\eta _1\eta _2\left( 1-\eta _3\right) \left( 1-\eta
_5\right) $, $p_{33}=\left( 1-\eta _1\right) \left( 1-\eta _2\right) \eta
_4\eta _6$, and $p_{44}=\left( 1-\eta _1\right) \left( 1-\eta _2\right)
\left( 1-\eta _4\right) \left( 1-\eta _6\right) $. Assume $p_1\geq p_2\geq
p_3\geq p_4$, $p_1>0$, and $A_i=p_i/p_1$, the optimal values of $F$ can be
obtained using Lagrangian Multipliers and the results are classified as

1. If $A_i>0$ for $i=2,3$, then $\eta _1=\eta _2=\left( 1+\sqrt{A_2}\right)
/\left( \sum_{i=1}^4\sqrt{A_i}\right) $, $\eta _3=\eta _5=1/\left( 1+\sqrt{%
A_2}\right) $, $\eta _4=\eta _6=\sqrt{A_3}/\left( \sqrt{A_3}+\sqrt{A_4}%
\right) $, and $F_{optimal}=\left( \sum_{i=1}^4A_i\right) /\left(
\sum_{i=1}^4\sqrt{A_i}\right) ^2$.

2. If $A_2>0$, $A_3=A_4=0$, then $\eta _1=\eta _2=\eta _4=\eta _6=1$, $\eta
_3=\eta _5=1/\left( 1+\sqrt{A_2}\right) $, and $F_{optimal}=\left(
1+A_2\right) /\left( 1+\sqrt{A_2}\right) ^2$.

3. If $A_2=A_3=A_4=0$, then $\eta _i=1$, and $F_{optimal}=1$.

\begin{figure}[!b]
\begin{center}
\vspace*{-0.5cm} \resizebox *{8cm}{4cm}{\includegraphics*{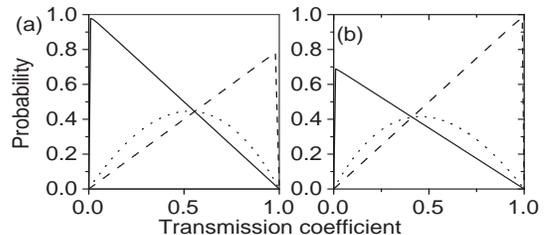}}
\end{center}
\par
\vspace*{-1.0cm}
\caption{The final successful probabilities $P$ (a) , $P^{\prime }$ (b)
versue transmission coefficient $\eta _1$. $A=10000$ (solid), $A=1$ (dot), $A=0.0001$ (dash).  (a) $k_1=0.8$; (b) $k_2=0.7$. }
\label{fig:3}
\end{figure}

So far we have described a scheme for preparing an arbitrary
polarization-entangled mixed state of two photons using variable beam splitters
and single mode optical fibres. In practical quantum information process, we
often use a special set of mixed states with the form $\rho =p\left| \psi
\right\rangle \left\langle \psi \right| +\left( 1-p\right) \left| \phi
\right\rangle \left\langle \phi \right| $, where $0\leq p\leq 1$ and $\left|
\psi \right\rangle $, $\left| \phi \right\rangle $ are arbitrary two-qubit
pure states. Our scheme can be simplified for this special mixed state.
Assume $\left| \psi \right\rangle =U_1\otimes V_1\left| \Phi \left( \alpha
\right) \right\rangle $, $\left| \phi \right\rangle =U_2\otimes V_2\left|
\Phi \left( \beta \right) \right\rangle $ with $\left| \Phi \left( \theta
\right) \right\rangle =\left( \cos \theta \left| HH\right\rangle +\sin
\theta \left| VV\right\rangle \right) $, and $0\leq \beta \leq \alpha \leq
\pi /4$, the experimental arrangement may be described by the schematic in
Fig. 2.

In Fig.2, VBS and SMOF perform same operations as those in Fig.1.
The distillation filters are used for entanglement transformation \cite
{Kwiat01} that is performed on location $\left| 1\right\rangle _A$ or $%
\left| 2\right\rangle _A$, depending on the initial state ($\left|
1\right\rangle _A$ corresponds to transformation $\left| \Phi \left( \beta \right)
\right\rangle \rightarrow \left| \Phi \left( \alpha
\right) \right\rangle$ and $\left| 2\right\rangle _A$ to $\left| \Phi \left( \alpha
\right) \right\rangle \rightarrow \left| \Phi \left( \beta \right)
\right\rangle $). The decoherence process yields different final
states for different initial states $\left| \Phi \left( \beta \right)
\right\rangle $ and $\left| \Phi \left( \alpha \right) \right\rangle $, 
\begin{eqnarray}
\rho &=&\frac 1P\left( k_1\eta _1\eta _2\left| \psi \right\rangle
\left\langle \psi \right| +\left( 1-\eta _1\right) \left( 1-\eta _2\right)
\left| \phi \right\rangle \left\langle \phi \right| \right),  \label{5} \\
\rho ^{\prime } &=&\frac 1{P^{\prime }}\left( \eta _1\eta _2\left|
\psi \right\rangle \left\langle \psi \right| +k_2\left( 1-\eta _1\right)
\left( 1-\eta _2\right) \left| \phi \right\rangle \left\langle \phi \right|
\right),  \nonumber
\end{eqnarray}
where $k_1=\sin ^2\beta /\sin ^2\alpha $ ($k_2=\cos ^2\alpha /\cos ^2\beta $
) \cite{Vidal99,zhang01} is the maximally feasible transformation
probability in experiment from $\left| \Phi \left( \beta \right)
\right\rangle $ to $\left| \Phi \left( \alpha \right) \right\rangle $ (from $%
\left| \Phi \left( \alpha \right) \right\rangle $ to $\left| \Phi \left(
\beta \right) \right\rangle $ ), $P=k_1\eta _1\eta _2+\left( 1-\eta
_1\right) \left( 1-\eta _2\right) $ and $P^{\prime }=\eta _1\eta
_2+k_2\left( 1-\eta _1\right) \left( 1-\eta _2\right) $ are the successful
probabilities to obtain $\rho $ and $\rho ^{\prime }$. The transmission
coefficients $\sqrt{\eta _1}$ and $\sqrt{\eta _2}$ satisfy 
conditions $\left( 1-\eta _1\right) \left( 1-\eta _2\right) =Ak_1\eta _1\eta
_2$ for $\rho $ and $k_2\left( 1-\eta _1\right) \left( 1-\eta _2\right)
=A\eta _1\eta _2$ for $\rho ^{\prime }$, where $A=\left( 1-p\right) /p$.
The optimization of $P$ and $P^{\prime }$ yields that 
\begin{eqnarray}
P &=&k_1\left( 1+A\right) /\left( 1+\sqrt{Ak_1}\right) ^2,  \label{6} \\
P^{\prime } &=&k_2\left( 1+A\right) /\left( \sqrt{k_2}+\sqrt{A}\right) ^2,
\nonumber
\end{eqnarray}
with $\eta _1=\eta _2=1/\left( 1+\sqrt{Ak_1}\right) $ for $\rho $ and $\eta
_1=\eta _2=\sqrt{k_2}/\left( \sqrt{k_2}+\sqrt{A}\right) $ for $\rho
^{\prime }$. Direct comparison between $P$ and $P^{\prime }$ shows
that if $0\leq p\leq \frac{k_1\left( 1-\sqrt{k_2}\right) ^2}{k_1\left( 1-%
\sqrt{k_2}\right) ^2+k_2\left( 1-\sqrt{k_1}\right) ^2}$, then $P^{\prime
}\leq P$ and $\left| \Phi \left( \beta \right) \right\rangle $ is chosen as
initial state. Otherwise $P\leq P^{\prime }$ and $\left| \Phi \left(
\alpha \right) \right\rangle $ is used. 

\begin{figure}[t]
\begin{center}
\vspace*{-0.5cm} \resizebox *{8cm}{4cm}{\includegraphics*{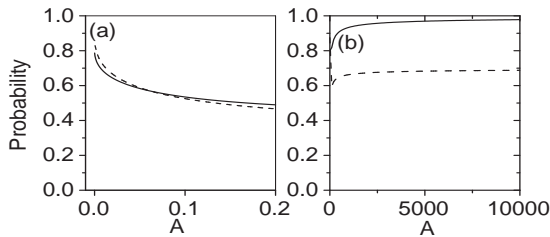}}
\end{center}
\par
\vspace*{-1.0cm}
\caption{The final successful probabilities $P$ (soild), $P^{\prime }$
(dash) versue the ratio $A$. $k_1=0.8$, $k_2=0.7$. (a) $0 \le A \le0.2$; (b) $0\le A\le10000$.}
\label{fig:4}
\end{figure}

In Fig.3, we plot the successful probabilities $P$, $P^{\prime }$ with
respect to the transmission coefficient $\eta _1$. The probability $P$ ($%
P^{\prime }$) reaches the maximum at certain $\eta _1$, as predicted by
Eq.(6). We notice that there exist fixed points $\left( \eta _1,P\right)
=\left( 1/\left( 1+k_1\right) ,k_1/\left( 1+k_1\right) \right) $ and $\left(
\eta _1,P^{\prime }\right) =\left( k_2/\left( 1+k_2\right) ,k_2/\left(
1+k_2\right) \right) $ for arbitrary parameter $A$. If the transmission
coefficient $\eta _1$ is selected at those points, the final successful
probabilities $P$, $P^{\prime }$ are constants, independent of the ratio
of two components $\left| \psi \right\rangle $ and $\left| \phi
\right\rangle $.

In Fig.4, we plot the optimal probabilities $P$, $P^{\prime }$ with
respect to the parameters $A$. As predicted by Eq.(6), $P$ $%
\rightarrow k_1$, $P^{\prime }\rightarrow 1$ as $A\rightarrow 0$; while $%
P\rightarrow 1$, $P^{\prime }\rightarrow k_2$ as $A\rightarrow \infty $.
These two cases correspond to pure final state $\rho $. We also notice that
there exist minimum values of $P$ ($P^{\prime }$) at $A=k_1 $($1/k_2$).

Fig.5 shows the change of success probability $P$($P^{\prime }$) as $\beta $ increases
from $0$ to $\alpha $. If $A\leq 1$, $P$ is always less than $%
P^{\prime }$ for any $\beta $ and initial state $\left| \Phi \left( \alpha \right)
\right\rangle $ is always used; while for $A>1$, $\left| \Phi \left( \beta \right)
\right\rangle $ may be used for large $\beta $.  

\begin{figure}[thb]
\begin{center}
\vspace*{-0.5cm} \resizebox *{8cm}{4cm}{\includegraphics*{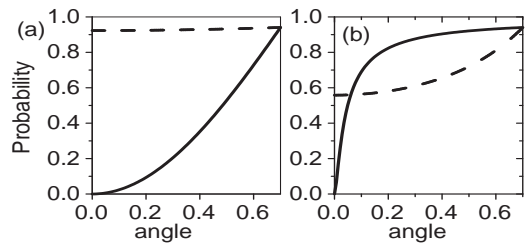}}
\end{center}
\par
\vspace*{-1.0cm}
\caption{The final successful probabilities $P$ (solid), $P^{\prime}$ (dash) versue the
angle $\beta $. $\alpha =0.7$. (a) $A=0.001$; (b) $A=1000$. }
\label{fig:5}
\end{figure}

In conclusion, we have described an experimental scheme for producing an
arbitrary polarization-entangled mixed state of two photons via controlled
location decoherence. The scheme uses only linear optical devices 
and single-mode optical fibers and may be feasible within current optical technology.
We believe the scheme may provide a useful mixed state entanglement source in the exploration of
various quantum information processing.

\end{document}